# Controlling heat and particle currents in nanodevices by quantum observation

Robert Biele[1*], César A. Rodríguez-Rosario[1,2*], Thomas Frauenheim[3], Angel Rubio[1,2*]

## Abstract

We demonstrate that in a standard thermo-electric nanodevice the current and heat flows are not only dictated by the temperature and potential gradient but also by the external action of a local quantum observer that controls the coherence of the device. Depending on how and where the observation takes place the direction of heat and particle currents can be independently controlled. In fact, we show that the current and heat flow in a quantum material can go against the natural temperature and voltage gradients. Dynamical quantum observation offers new possibilities for the control of quantum transport far beyond classical thermal reservoirs. Through the concept of local projections, we illustrate how we can create and directionality control the injection of currents (electronic and heat) in nanodevices. This scheme provides novel strategies to construct quantum devices with application in thermoelectrics, spintronic injection, phononics, and sensing among others. In particular, highly efficient and selective spin injection might be achieved by local spin projection techniques.

## Introduction

Classical non-equilibrium thermodynamics was developed to understand the flow of particles and energy between multiple heat and particle reservoirs[1]. The best-known example is Clausius' formulation of the second law of thermodynamics stating that heat cannot flow from a cold bath to a hot one[2]. This is firmly based on the assumption that a macroscopic body in equilibrium is characterized by a single parameter: its temperature. When two objects with different temperatures are brought in contact, heat will flow from the hotter to the colder one. In macroscopic objects, the observation of this process does not influence the flow of energy and particle between them. However, in quantum materials, thermodynamical concepts have to be revisited[3,4]. There, states of matter can be set into a coherent superposition, such as the famous

[1] Nano-Bio Spectroscopy Group and ETSF, Department of Materials Science, Universidad del País Vasco UPV/EHU, E-20018 San Sebastián, Spain.
[2] Max Planck Institute for the Structure and Dynamics of Matter Hamburg and Center for Free-Electron Laser Science & Department of Physics, University of Hamburg, Luruper Chaussee 149, 22761, Germany.
[3] Bremen Center for Computational Materials Science, Bremen 28359, Germany.
* Correspondence and requests for materials should be addressed to R.B. (r.biele02@gmail.com, +34 943 01 8497), C.A.R.R. (crodrig@mpsd.mpg.de, +49 40 8998 6643) and to A.R. (angel.rubio@mpsd.mpg.de, +49 40 8998 6550).

Schrödinger's cat. When a classical observer measures a nanoscale system, this interaction destroys most of the coherence inside the system and alters its dynamical response.[5]

However, if a quantum observer instead acts only locally, the system quantum coherence changes continuously and dynamically. Depending on how strong and where these local quantum observations are performed, novel and surprising quantum transport phenomena might arise. This effect of local quantum observation is quite general and covers a wide range of scenarios, it can come for example from a quantum observer of the kind discussed in [10], continuous quantum measurements[6] or the effect of quantum observation can come from local electron-phonon coupling to another quantum system. Decoherence[7], for instance, has been shown to affect the dynamical evolution and decoupling from the environment via the Quantum Zeno effect[8,9], the rate of cooling[10] and the efficiency of energy transport in molecular devices[11,12] and biological systems[13,14,15]. Here, to include the role of the quantum observer into a complete consistent thermodynamic formalism, we will use an external quantum bath that changes quantum coherences in the basis defined by the observer. In contrast to standard reservoirs in classical thermodynamics, described to be in an ensemble thermal state, our quantum coherent bath has no temperature parameter that can be associated to it[16]. This offers further possibilities for the control of quantum transport far beyond classical reservoirs. The present work focuses on the role of this particular quantum observer and the non-equilibrium thermodynamic[17,18,19] implications it has on both particle and energy transport through thermo-electric nanodevices[20,21,22]. Surprisingly, we find that the quantum observer can emergently[23] control heat and particle current, both in direction and strength. Therefore, incorporation of the concept of a quantum observer can lead to novel strategies to construct quantum devices with application in thermoelectrics, spintronic injection, phononics, and sensing, to just name a few. For instance, controlling the direction and strength of the heat flow will improve the figure of merit of thermoelectric devices by decreasing the thermal conductivity and outperform current technologies. In addition, the projection of only one component of the spin might open the path towards an efficient polarized spin injection strategy.

**Results**

To illustrate the new phenomena induced by the quantum observer, we consider the general and standard transport device shown in Fig. 1. Without loss of generality, this device is modeled by a simple tight-binding Hamiltonian with uniform on-site energy and nearest-neighbor hopping of 0.5 eV. For simplicity, we assume a single electron to be present in the device, though the results would not change if many electrons were active. This configuration can easily be realized experimentally in silicon heterostructures[24], with cold atoms[25] or in graphene[26]. Our findings are universal and apply to more general nanodevices: The geometry of the device modifies the absolute value of the current flows but not the effect itself. The nine leftmost and rightmost sites are connected to macroscopic thermal baths at different temperature. As a result, energy and particles will flow equally through the two branches. For the sake of simplicity in all this work we set the external voltage gradient between the reservoirs to zero; all effects in heat and particle flows come from the thermal gradient and the quantum observer alone. Adding an external bias voltage would introduce another source of current but would not modify the conclusions concerning the impact of the quantum observer in the dynamics of the energy and particle flow in the nanodevice. While in this configuration energy can be exchanged via the baths, particle current is then corralled inside the device. At steady state no particle current is present, however, constant heat flows from hot to cold according to the second law of thermodynamics.

While the coupling to the two thermal baths is modeled by a standard master equation[27], the quantum observer acts on one site changing the quantum coherence only as in the double-slit experiment[28]. The quantum observer is assumed to be in a pure quantum state rather than in an ensemble temperature-state as thermal baths are and therefore cannot have a temperature associated to it. We are interested here in understanding how the particle and heat currents at steady state change when the quantum observation acts on a specific site, as indicated by the eye in Fig. 2a. For that specific case, Fig. 2b shows the heat current at steady state through the device as a function of the coupling strength to the quantum observer $\gamma_D$ and the amplitude of the applied thermal gradient $\Delta T$. A positive current (red) indicates a flow from left to right. While the upper curved surface shows the energy current in the upper branch of the device, $j_{\text{up}}^h$, the flat contour-plot below corresponds to the energy flow in the lower branch, $j_{\text{down}}^h$. We emphasize that this contour plot is a projection onto the plane, hence only the color gradient indicates its strength. When the quantum observer measures the system at site labeled $\alpha$, the energy current increase in the natural direction of the thermal gradient. However, even in the absence of a thermal gradient, the effect of the observation is to create a quantum heat flow from left to right. Remarkable, this local observation induces also a particle current in the upper branch from left to right, as can be clearly seen in Fig. 2c. As we are at the steady state, the corresponding particle current in the lower branch is exactly the opposite of it. As a result of the local action of the observer at site $\alpha$, an unexpected particle ring-current is induced in the device that flows in clockwise direction. By symmetry, a similar measurement on site $\gamma$ would give a counter-clockwise ring-current.

We now consider how this situation is modified when we change the observation site from $\alpha$ to $\beta$, indicated in Fig. 2d. Most surprisingly, now the heat flows change direction as seen in Fig. 2e: When no observation is performed (triangle 1), heat goes from the hot to the cold reservoirs in the upper and lower branches as expected. However, beyond a certain observer coupling strength $\gamma_D$ (triangle 3), the *energy moves in both branches against the thermal gradient*, that is, heat goes from the cold to the hot bath. Additionally, for intermediate coupling strength $\gamma_D$ (triangle 2) we observe energy ring-currents in counter-clockwise direction. Interestingly, the observation induces now a *counter*-clockwise particle ring-current, as can be seen in Fig. 2f. This is a consequence of the localization of the electronic state induced by the local observation. As the electronic density in our model is just four times larger in the leads than in the branches, the quantum observer acting on site $\beta$ pulls the electron out of the right lead pushing it towards the left lead. An electronic current starts to flow in the upper branch from right to left. When instead the observation is performed close to the left lead, the particle flows in the opposite direction.

To show that this new effect is general and gives rise to even more interesting applications and quantum phenomena in more complex nanoscale systems, we next consider another device similar to the famous 'Feynman's ratchet'[29]. Quantum ratchets are transport devices driven by thermal or quantum fluctuations that have been widely studied to control the flow of particles and heat[30,31]. In order to model such a ratchet, we introduce the spatial asymmetry by changing the on-site energy levels on the parallel branches of our device shown in Fig. 3a. In the upper branch (sites $\alpha - \beta$), the on-site energies are increased steadily by 10% as we move from left to right, which is graphically indicated by the size of the spheres. In the lower branch instead, the on-site energies increase from right to left in equal proportion. Therefore, we have created a quantum ratchet by adding two rectifiers on each branch in opposite direction. This could be experimentally realizable with techniques developed for constructing quantum ratchets in graphene[26], atom traps[32] or for molecular junctions[33]. This configuration is chosen here to illustrate the differences of making a quantum observation in the top or bottom branches by breaking the top–down symmetry of the original device of Fig. 1. First, Fig. 3c shows that even without an observation a particle current flows in clockwise direction. This is because the system acts similar as a "salmon ladder" for our chosen working temperature[3]. In the upper branch for instance, there is a small probability of the particle to hop from the left side each step up the ladder due to thermal fluctuations, but there is a very low probability for the particle on the right side to climb up the big drop of the ladder to then go down the ladder. Most interestingly, by adding a quantum observation on site $\beta$, the particle current decreases and later goes against the natural direction of the ratchet. When the observation is instead performed at site $\delta$, in the bottom branch, the particle current is always in the preferred direction of the ratchet, but it can be significantly increased with stronger observation strength (See Supplemental Material). This effect can be explained by similar arguments as before for the flat geometry and is additionally illustrated as an animation in the Supplemental Material. The energy currents, as illustrated in Fig. 3b, have similar regimes as discussed before.

## Discussion

We highlight that the particle current changes its direction under very weak coupling to the quantum observer. The quantum ratchet device could hence be used for sensing applications with nanoscale resolution observing weak quantum effects. Since the quantum observer is in the thermodynamics limit, this behavior can be called emergent[23]. Having heat flowing from a cold bath to a hotter one seems to be an apparent violation of Clausius' formulation of the second law of thermodynamics, but it is not. In order to understand this, we examine the situation in terms of basic non-equilibrium thermodynamic entropy concepts[2]. The general entropy production rate equation[34] can be written as

$$\dot{S} = \Phi + P, \quad (1)$$

where $S$ is the total entropy of the system, $\Phi$ is the net entropy flow into the system and $P$ is the entropy production due to irreversibility inside the system that is always positive. For a system in the steady state between two temperatures, $T_H$ and $T_C$, $\dot{S}$ should be zero and

$$\Phi = \Phi_H + \Phi_C = -\frac{\dot{Q}_H}{T_H} - \frac{\dot{Q}_C}{T_C} = -P. \quad (2)$$

Here, $\dot{Q}_H$ and $\dot{Q}_C$ denote the heat flows to the hot and cold reservoirs, respectively. This leads to the following equation

$$\frac{\dot{Q}_H}{T_H} + \frac{\dot{Q}_C}{T_C} = P \geq 0. \quad (3)$$

In the case that the only energy exchange happens via the two baths, the continuity equation, $-\dot{Q}_H = \dot{Q}_C$, leads to the familiar concept that heat goes from hot to cold reservoirs. However, by introducing the coupling to the quantum observer, it is indeed possible to have heat flowing in more complicated ways, even against the thermal gradient. This is because the observer can create an energy flow even if it does not have a temperature associated to it. The new energy source changes the continuity equation to $\dot{Q}_H + \dot{Q}_C + \dot{Q}_D = 0$, where $\dot{Q}_D$ is a purely quantum coherent heat flow introduced by the observation process. Previous seminal work on projective measurements showed that the fluctuation theorems still holds and that a complete set of measurements does not alter the forward and backwards probability ratios[20]. In our case, we consider a quantum observer that operates continuously over time only on a single site instead. As our observer acts on a local subspace of the system, it breaks its symmetry and hence introduces particle ring currents and allows controlling the energy currents inside quantum materials. This furthers the idea of the importance of local quantum observations and where they are made as an important control parameter. This quantum observer does not add a new entropy flow to Eq. (2) (as shown in Methods), but changes the particle and heat flows. The observer effectively acts as a quantum source of heat by changing the entropy production $P$ directly and breaking the symmetry of the system. This creates changes in the electronic steady state in quantum materials in analogy to dissipative structures[34]. Classical dissipative structures have been used to explain non-equilibrium systems such as living organisms or hurricanes[35]. Like in classical dissipative structures the increase in entropy production, due to the quantum measurement, leads to the observed reversal of entropy flows in our device (See SI).

In conclusion the results presented here indicate that particle and energy currents in nanoscale systems can be created and controlled by the mere presence of a quantum observer. The present work can be further connected to the concept of entropy-driven organized dynamical systems, also known as dissipative structures[34]. However, in our nanodevice, the increase in entropy production comes from a purely quantum mechanical source. We have shown that understanding and controlling the role of a quantum observer can lead to advances in thermoelectric materials. Additionally, a quantum observer can be used for novel ways of creating and controlling energy and particle transport that can have applications for spintronic injection, quantum phononics, and sensing. For example, a local observer can act as a magnetic memory writer that projects out one of the spin components, giving rise to a complete spin-polarized current even in the absence of spin-orbit coupling. It also highlights the role of an observer in quantum devices: While in the famous Schrödinger's cat paradox the coherent state is destroyed directly via the observation by a classical object, here we have shown that a local quantum observer changes coherence locally and dynamically modifying the heat and electronic transport behavior of the device. Quantum measurements offer new possibilities of thermodynamic control for quantum transport[36] far beyond classical thermal reservoirs. Additionally, the quantum observer might also be used to control the local temperature distribution of quantum devices[37,38].

## Methods

Our device in Fig.1 is modeled by the simple time-independent Hamiltonian

$$H = \sum_i \epsilon_i c_i^\dagger c_i - \mathcal{T} \sum_{\langle i,j \rangle} (c_i^\dagger c_j + c_j^\dagger c_i), \quad (4)$$

where $\langle i,j \rangle$ is the nearest neighbor index and $c_i^\dagger$ creates an electron at site $i$. For simplicity, only one electron is present in our study. To introduce a thermal imbalance $\Delta T$ in the transport device, we connect the nine leftmost and rightmost atoms to thermal baths kept at temperature $T_H$ and $T_C$ respectively. These temperatures are chosen around an average temperature $T_E$ such that $T_{H,C} = T_E \pm \Delta T$. The state of the system evolves according to a standard Markovian master equation[27] (All equations are written in atomic units):

$$\frac{d\rho}{dt} = -i[H,\rho] + \sum_{\alpha=H,C,D} (K_\alpha \rho S_\alpha + S_\alpha \rho K_\alpha - S_\alpha K_\alpha \rho - \rho K_\alpha S_\alpha)$$
$$= -i[H,\rho] + \sum_{\alpha=H,C,D} \mathbb{L}_\alpha[\rho], \quad (5)$$

where $K_{H,C} = \lambda^2 \int_{-\infty}^{\infty} C_{H,C}(\tau) e^{-iH\tau} S_{H,C} e^{iH\tau} d\tau$ describes the influence of the two thermal baths onto the system. We point out that the master equation (5) has the same steady-state solution as a second-order non-Markovian master equation[27]. The bath-correlation function $C$ for the hot and cold thermal baths can be derived from first principle by assuming the electronic system is in contact with the radiation inside a cavity[27,39]

$$C_{H,C}(\tau) = \frac{1}{2\epsilon_0 \pi} \int_0^{\omega_c} d\omega \{[n_B(\omega, T_{H,C}) + 1]e^{-i\omega\tau} + n_B(\omega, T_{H,C})e^{i\omega\tau}\}, \quad (6)$$

where $n_B$ is the Bose-Einstein distributing function and $\omega_c$ a cut-off frequency due to the dipole approximation. Furthermore, the coupling operator $S$ in Eq. (5) is given by

$$S_{H,C} = -\boldsymbol{u} \cdot \boldsymbol{r}\, M_{H,C}(\boldsymbol{r}), \quad (7)$$

where $e$ is the electron charge, $\boldsymbol{u}$ the polarization direction of the modes in the cavity (all three spatial coordinates) and $M_{H,C}(\boldsymbol{r})$ is a mask function that is either one for $\boldsymbol{r}$ in the hot and cold region or otherwise zero. This will model the local coupling at the left or right ends of the device. This choice of the bath-correlation function (6) and the coupling operators (7) can be derived from first-principles[39] and ensures that the system relaxes towards thermal equilibrium in the case of zero temperature gradient ($T_H = T_C$). This is indeed the case in our simulations. In order to account for the local quantum measurement within the same thermodynamic formalism, we consider $K_D = \frac{\gamma_D^2}{2}|k\rangle\langle k|$ and $S_D = |k\rangle\langle k|$, where $k$ indicates the site where the observer is acting with observation strength $\gamma_D$. In this way, the quantum observer is unital and changes quantum coherences to site $k$ continuously and dynamically. This kind of projection does not freeze the dynamics, like in the quantum Zeno Effect, but does affect the final steady state and thus the macroscopic measurable thermodynamic flows. This effect of local observation can come from a quantum observer of the kind discussed in [10] or for example from electron-phonon coupling. For the latter case the form of the operators $S_D$, $K_D$ and the master equation (5) for the quantum observer can be derived by considering random white-noise fluctuations of the local energy density and averaging over many realizations of the noise. In this way one derives at the thermodynamic formalism applied in this work.

In order to study the energy flows through our transport device, we define the local energy-density operator $h_i = -\frac{1}{2}\mathcal{T}\sum_{\langle j\rangle}(c_i^\dagger c_j + c_j^\dagger c_i) + \epsilon_i c_i^\dagger c_i$ where the summation is over the neighboring sites. Then, the continuity equation, $-\frac{dh_i}{dt} = -i[H, h_i]$, is used to define the energy-current operator[40] that describes the heat flow in the middle of the upper branch, $j_{up}^h$,

$$j_{up}^h = \frac{i}{2}\left[\epsilon_3 c_3^\dagger c_3 - \epsilon_4 c_4^\dagger c_4, -\mathcal{T}(c_3^\dagger c_4 + c_4^\dagger c_3)\right]$$
$$+ \frac{i\mathcal{T}^2}{2}\left([c_3^\dagger c_4 + c_4^\dagger c_3, c_4^\dagger c_5 + c_5^\dagger c_4] + [c_2^\dagger c_3 + c_3^\dagger c_2, c_3^\dagger c_4 + c_4^\dagger c_3]\right), \quad (8)$$

where sites are numbered from left to right with $\alpha = 1$ and $\beta = 5$. Note, that the energy current is not an observable as it depends on the choice of the local energy-density operator. The energy current in the lower branch, $j_{down}^h$, can be defined in a similar way. Moreover, by considering the electronic density the particle current $j_{up}^p$ in the upper branch is defined in a similar way. Those operators have been used to calculate the expectation values at the steady state ($\dot{\rho} = 0$) reached in the long-time limit of the dynamics.

As parameters for the model, we chose $\epsilon_0 = 1$ eV $= 0.036$ a.u., $T = \epsilon_0/2$, $\lambda/\sqrt{\epsilon_0} = 0.2$ and $k_B T_E = 0.008$ a.u. We studied further ranges of $T_E$ but we will present this in a future publication. We considered devices in two configurations. In the flat configuration, as seen in Fig. 1, all sites have the same on-site energy of 0 eV. The second configuration studied is the quantum ratchet seen in Fig. 3a. For that, the sites of the right and left leads have energy 0 eV. Sites $\alpha$ to $\beta$ form a ladder of equal steps in energy of 0.1 eV, such that $\epsilon_\alpha = 0.1$ eV, ..., $\epsilon_\beta = 0.5$ eV. Sites $\gamma$ to $\delta$ form a ladder of equal steps of 0.1 eV in the opposite direction, such that $\epsilon_\gamma = 0.5$ eV and $\epsilon_\delta = 0.1$ eV. Because these rectifiers are anti-parallel, they create a preferred clockwise particle flow as explained in the main text.

**A quantum observer has zero entropy flow.** Examining the flows due to the local observation shows that the quantum observer does not add a new entropy flow to the system in contrast to a standard thermodynamic heat bath, and therefore the quantum observer cannot be described as a heat bath at a certain temperature. For the coupling described by the master equation (5), the entropy flow into the system due to the local observer can be written as[16]

$$\Phi_D = -\text{Tr}[\mathbb{L}_D[\rho] \ln \sigma_D]. \qquad (9)$$

Here, $\sigma_D$ is the stationary state of the local quantum observation at site $k$, $\sigma_D = |k\rangle\langle k|$, and $\mathbb{L}_D$ is defined in Eq. (5) and leads to

$$\mathbb{L}_D[\rho] = \gamma_D^2 [2|k\rangle\langle k|\rho|k\rangle\langle k| - |k\rangle\langle k|\rho - \rho|k\rangle\langle k|]. \qquad (10)$$

Inserting the former equation into Eq. (9) shows that the entropy flux due to the quantum observer is zero. This means that a quantum observer changes the energy flow in the system directly by introducing work in the system, without having an entropy flow connected with it. Additionally, one can see that Eq. (10) leaves the identity invariant, $\mathbb{L}_D[\mathcal{I}] = 0$, hence the dynamics is unital.

## Acknowledgments
We acknowledge financial support from the European Research Council (ERC-2015-AdG-694097), Grupos Consolidados (IT578-13) and European Union's Horizon 2020 Research and Innovation program under Grant Agreements no. 676580 (NOMAD) and the Marie Sklodowska-Curie Individual fellowships H2020-MSCA-IF-2015 grant no. 706890. We acknowledge Florian Eich, Fabio Covito, Roberto D'Agosta and Peter Hänggi for interesting discussions and helpful comments. R. B. acknowledges the financial support of the Ministerio de Educacion, Cultura y Deporte (Grant No. FPU12/01576).

## Author contributions
All authors have contributed extensively to this work. R.B. and C.A.R.R. provided theoretical support and performed the simulation and data analysis. A.R. and T.F. conceived, designed and led the research.

## Supplementary information is available at npj Quantum Material's website.

## Figures

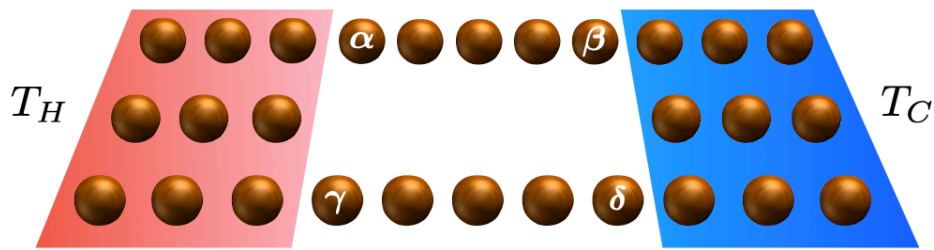

**Figure 1 | Sketch of the thermo-electric transport nanodevice studied here.** The nine leftmost and rightmost atoms are connected to thermal baths at temperature $T_H$ and $T_C$, respectively. Due to this temperature imbalance heat flows from the hot to the cold side equally through the two identical branches. This device will be used to study physical phenomena arising from local quantum observations. We tested similar devices with different geometries and lengths of the two branches. The conclusions drawn with the present configuration remain valid.

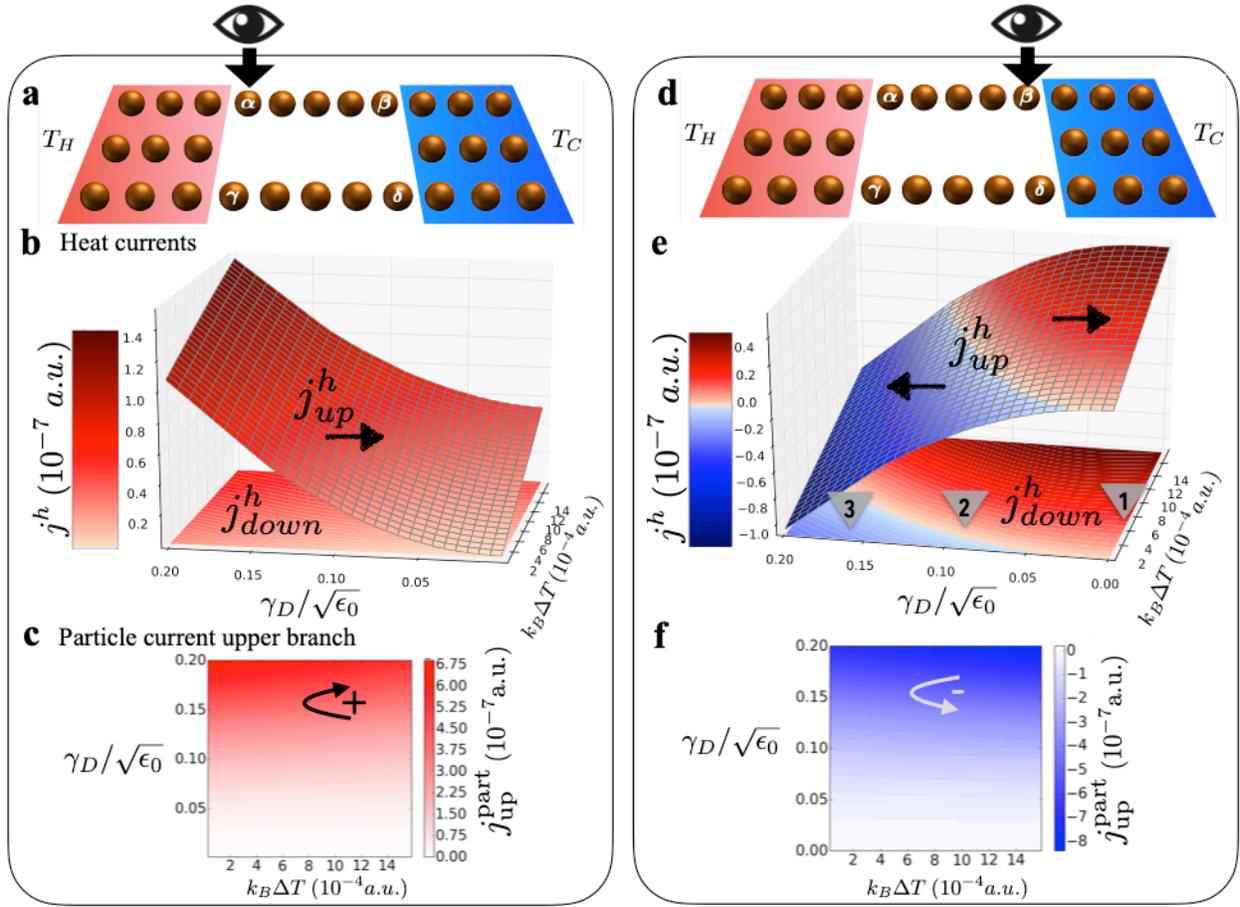

**Figure 2 | Particle and energy currents in the steady state.** In **a**–**c** the observation is performed at site $\alpha$, while in **d**–**f** on site $\beta$. **b** and **e** show the heat current in the upper, $j^h_{up}$, and lower branch, $j^h_{down}$. In order to allow comparison of the both, the upper current is plotted as a curved surface, while the lower energy current has been projected onto the plane below. We emphasize that this contour plot is a projection onto the plane, only the color gradient indicates its strength. A positive current (red) represents heat flowing from left to the right. In **c** and **f** the particle current in the upper branch is shown. A positive current indicates that a particle ring-current is flowing clockwise. The triangles labeled 1, 2 and 3 in **e** mark regions where the energy flow is to the right in both branches, in different directions in each branch, and to the left in both branches, respectively. A temperature gradient of $10^{-3}$ a.u. corresponds to around 300 K, and a particle current of $3 \times 10^{-7}$ a.u. corresponds to 2 nA.

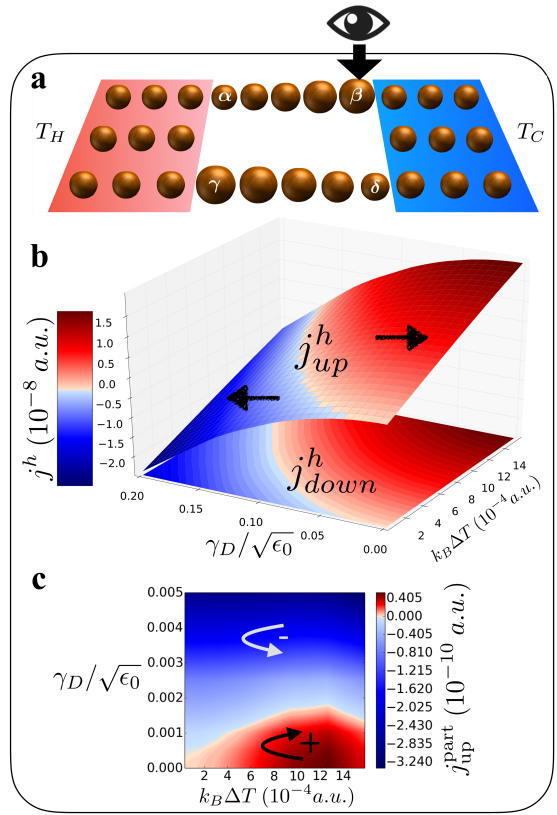

**Figure 3 | Influence of a quantum observation on the thermo-electric flows in a quantum ratchet**. (**b**) Change of the energy currents in the top ($j^h_{up}$) and bottom ($j^h_{down}$) branches due to the influence of a local measurement on site $\beta$. (**c**) The particle current in the upper branch can be seen. It is striking in **c** that the particle ring-current can change direction as a function both of $\gamma_D$ and $\Delta T$.